\documentclass[10pt]{article}
\usepackage[english]{babel}
\usepackage{graphics}
\usepackage{graphicx}
\usepackage{amsmath}
\usepackage{amsfonts}
\usepackage{amssymb}%
\begin{document}
\def\beq{\begin{equation}}
\def\eeq{\end{equation}}
\def\bear{\begin{eqnarray}}
\def\ear{\end{eqnarray}}
\newcommand{\Eq}[1]{Eq.\,(\ref{#1})}
\newcommand{\sect}[1]{Sec.\,#1}
\newcommand{\Ref}[1]{Ref.\,\cite{#1}}
\newcommand{\Refs}[1]{Refs.\,\cite{#1}}
\def\s{{\,\rm s}}
\def\g{{\,\rm g}}
\def\eV{\,{\rm eV}}
\def\keV{\,{\rm keV}}
\def\MeV{\,{\rm MeV}}
\def\GeV{\,{\rm GeV}}
\def\TeV{\,{\rm TeV}}
\def\sv{\left<\sigma v\right>}
\def\({\left(}
\def\){\right)}
\def\cm{{\,\rm cm}}
\def\K{{\,\rm K}}
\title{Primordial heavy elements in composite dark matter models}
\author{Maxim Yu. Khlopov$^{1,2,3}$\thanks{Maxim.Khlopov@roma1.infn.it}}

\date{}
\maketitle

\begin{center}
{\sl\small $^{1}$ Centre for Cosmoparticle Physics "Cosmion"\\
Moscow, 125047,
Miusskaya pl. 4 \\
$^{2}$ Moscow State Engineering Physics Institute, \\
Kashirskoe Sh., 31, Moscow 115409, Russia, and \\
$^{3}$ APC laboratory 10, rue Alice Domon et L\'eonie Duquet \\75205 Paris
Cedex 13, France}
\end{center}

\begin{abstract}
A widely accepted viewpoint is to consider candidates for cosmological dark matter as neutral and weakly interacting particles, as well as to consider only light elements in the pregalactic chemical composition. It is shown that stable charged leptons and quarks can exist and, hidden in elusive atoms, play the role of dark matter. The inevitable consequence of realistic scenarios with such composite atom-like dark matter is existence of significant or even dominant fraction of "atoms", binding heavy -2 charged particles and $^4He$ nuclei. Being $\alpha$ particles with shielded electric charge, such atoms catalyse a new path of nuclear transformations in the period of Big Bang Nucleosynthesis, which result in primordial heavy elements. The arguments are given, why such scenario escapes immediate contradiction with observations and challenges search for heavy stable charged particles in cosmic rays and at accelerators.
\end{abstract}

\section{Introduction}
The problem of existence of new particles is among the most
important in the modern high energy physics. This problem can acquire
cosmological impact, if the particles are longliving. Then
their presence in the Universe can be traced in the astrophysical data, in particular, in the data on normal baryonic matter (see e.g. \cite{book,Cosmoarcheology,Bled06} for review and references). 

The theory of Big Bang Nucleosynthesis \index{Big Bang Nucleosynthesis} and the observational data on chemical element abundance is an important tool in studies of possible properties of new particles. It is widely known that Standard Big Bang Nucleosynthesis gives rise to formation of light elements only and its theory predicts negligible pregalactic abundance of elements, heavier than lithium. Here we show that this point can change drastically, if there exist stable charged leptons and/or quarks.

Recently several elementary particle frames for heavy
stable charged particles were considered: (a) A heavy quark \index{quark} and
heavy neutral lepton \index{lepton} (neutrino \index{neutrino} with mass above half the Z-Boson
mass) of fourth generation \cite{N,Legonkov}; which can avoid
experimental constraints \cite{Q,Okun} and form composite dark
matter species \cite{I,lom,KPS06,Khlopov:2006dk}; (b) A Glashow's ``Sinister''
heavy tera-quark \index{tera-quark} $U$ and tera-electron \index{tera-electron} $E$, which can form a tower
of tera-hadronic and tera-atomic bound states with ``tera-helium
atoms'' $(UUUEE)$ considered as dominant dark matter
\cite{Glashow,Fargion:2005xz}. (c) AC-leptons, predicted
in the extension \cite{5} of standard model, based on the approach
of almost-commutative geometry \cite{bookAC}, can form evanescent
AC-atoms, playing the role of dark matter \index{dark matter}
\cite{5,FKS,Khlopov:2006uv}. 

Finally, it was recently shown in \cite{KK} that an elegant solution is possible in the
framework of walking technicolor \index{technicolor} models
\cite{Sannino:2004qp,Hong:2004td,Dietrich:2005jn,Dietrich:2005wk,Gudnason:2006ug,Gudnason:2006yj}
and can be realized without an {\it ad hoc} assumption on charged
particle excess, made in the approaches (a)-(c).

In all these models, predicting stable charged particles, the
particles escape experimental discovery, because they are hidden
in elusive atoms, maintaining dark matter of the modern Universe.
It offers new solution for the physical nature of the cosmological
dark matter. Influence on the primordial chemical composition and appearance of primordial heavy elements \index{primordial heavy elements} is an inevitable consequence of this solution.

Indeed, it turned out that the necessary condition for the considered scenario, avoiding anomalous 
isotopes overproduction, is absence of stable particles with charge -1, so that stable negatively charged
particles should only have charge -2.  After it is formed in Big Bang Nucleosynthesis, $^4He$ screens the
$A^{--}$ charged particles in composite $(^4He^{++}A^{--})$ {\it O-helium} 
``atoms''. These neutral primordial nuclear interacting objects
contribute the modern dark matter density and play the role of a
nontrivial form of strongly interacting dark matter \index{strongly interacting dark matter}
\cite{Starkman,McGuire:2001qj}. The active influence of this type
of dark matter on nuclear transformations seems to be incompatible
with the expected dark matter properties. However, it turns out
that the considered scenario is not easily ruled out \cite{FKS,I,KK}
and challenges the experimental search for various forms of O-helium and its
charged constituents.
\section{Charged components of composite dark matter}
\subsection{Charged tera-particles}
In Glashow's  "Sinister" $SU(3)_c \times SU(2) \times SU(2)'
\times U(1)$ gauge model \cite{Glashow} three heavy generations of
tera-fermions are related with the light fermions by $CP'$
transformation linking light fermions to charge conjugates of
their heavy partners and vice versa. $CP'$ symmetry breaking makes
tera-fermions much heavier than their light partners. Tera-fermion
mass pattern is the same as for light generations, but all the
masses are multiplied by the same factor $\sim 10^6$. It gives the
masses of lightest heavy particles in {\it tera}-eV (TeV) range,
explaining their name.  
Strict conservation of $F = (B-L) - (B' - L')$ prevents mixing of
charged tera-fermions with light quarks and leptons. Tera-fermions
are sterile relative to $SU(2)$ electroweak interaction, and do
not contribute into standard model parameters.
 In such realization the new heavy neutrinos ($N_i$)
  acquire large masses and their
mixing with light neutrinos $\nu$ provides a "see-saw" mechanism
of light neutrino Dirac mass generation. Here in a Sinister model
the heavy neutrino is unstable. On the contrary in this scheme tera-electron
$E^-$ is the lightest heavy fermion and it is absolutely stable.

Since the lightest quark $U$ of heavy generation does not mix with
quarks of 3 light generation, it can decay only to heavy
generation leptons owing to GUT-type interactions, what makes it
sufficiently long living. If its lifetime exceeds the age of the
Universe, primordial $U$-quark hadrons as well as heavy leptons
$E^-$ 
should be present in the modern matter.

Glashow's "Sinister" scenario \cite{Glashow} took into account
that
 very heavy quarks $Q$ (or antiquarks $\bar Q$) can form bound states with other heavy quarks
 (or antiquarks) due to their Coulomb-like QCD attraction, and the binding energy of these states
 substantially exceeds the binding energy of QCD confinement.
Then $(QQq)$ and $(QQQ)$ baryons can exist.

According to  \cite{Glashow} primordial heavy quark $U$ and heavy
electron $E$ are stable and
may form a neutral most probable and stable (while being
evanescent) $(UUUEE)$ "atom"
with $(UUU)$ hadron as nucleus and two $E^-$s as "electrons". The
tera gas of such "atoms" seemed an ideal candidate for a very new
and fascinating dark matter;
because of their peculiar WIMP-like interaction with matter they
might also rule the stages of gravitational clustering in early
matter dominated epochs, creating first gravity seeds for galaxy
formation. \subsection{Stable AC leptons from almost commutative
geometry} The AC-model \index{AC-model} \cite{5} appeared as realistic elementary
particle model, based on the specific approach of \cite{bookAC} to
unify general relativity, quantum mechanics and gauge symmetry.

This realization naturally embeds the Standard model, both
reproducing its gauge symmetry and Higgs mechanism, but to be
realistic, it should go beyond the standard model and offer
candidates for dark matter. Postulates of noncommutative geometry
put severe constraints on the gauge symmetry group, excluding in this approach superstring, supersymmetric and GUT extensions.
The AC-model \cite{5} extends the fermion content of the Standard
model by two heavy particles with opposite electromagnetic and
Z-boson charges. Having no other gauge charges of Standard model,
these particles (AC-fermions) behave as heavy stable leptons with
charges $-2e$ and $+2e$, called $A$ and $C$, respectively.
AC-fermions are sterile relative to $SU(2)$ electro-weak
interaction, and do not contribute to the standard model
parameters. The mass of AC-fermions is originated from
noncommutative geometry of the internal space (thus being much
less than the Planck scale) and is not related to the Higgs
mechanism. The lower limit for this mass follows from absence of
new charged leptons \index{charged leptons} in LEP. In the absence of AC-fermion
mixing with light fermions, AC-fermions can be absolutely stable.

The mechanism of baryosynthesis in the AC model is not clear, therefore the AC-lepton excess
was postulated in \cite{5,FKS,Khlopov:2006uv} to saturate the
modern CDM density (similar to the approach of sinister model).
 Primordial excessive negatively charged $A^{--}$ and positively
charged $C^{++}$ form a neutral most probable and stable (while
being evanescent) $(AC)$ "atom", the AC-gas of such "atoms" being
a candidate for dark matter,
accompanied by small ($\sim 10^{-8}$) fraction of $^4HeA^{--}$ atoms
called in this case OLe-helium \index{OLe-helium} \cite{5,FKS,Khlopov:2006uv,Khlopov:2006dk}.

\subsection{Stable pieces of 4th generation matter} Precision data
on Standard model parameters admit \cite{Okun} the existence of 4th generation
\index{4th generation}, if 4th neutrino \index{4th neutrino} ($N$) has mass about 50 GeV, while
masses of other 4th generation particles are close to their
experimental lower limits, being $>100$GeV for charged lepton \index{charged lepton}
($E$) and $>300$GeV for 4th generation $U$ and $D$ quarks
\cite{CDF}. 

4th generation can follow from heterotic string \index{heterotic string} phenomenology and
its difference from the three known light generations can be
explained by a new conserved charge, possessed only by
its quarks and leptons \cite{N,Q,I}. Strict conservation of this charge makes the
lightest particle of the 4th family (4th neutrino $N$) absolutely
stable, while the lightest quark must be sufficiently long living
\cite{Q,I}. The lifetime of $U$ can exceed the age of the
Universe, as it was revealed in \cite{Q,I} for $m_U<m_D$.

$U$-quark can form lightest $(Uud)$ baryon and $(U \bar u)$ meson.
The corresponding antiparticles are formed by $\bar U$ with light
quarks and antiquarks. Owing to large chromo-Coulomb binding
energy ($\propto \alpha_{c}^2 \cdot m_U$, where $\alpha_{c}$ is
the QCD constant) stable double and triple $U$ bound states
$(UUq)$, $(UUU)$ and their antiparticles $(\bar U \bar U \bar u)$,
$(\bar U \bar U \bar U)$ can exist
\cite{Q,Glashow,Fargion:2005xz}. Formation of these double and
triple states in particle interactions at accelerators and in
cosmic rays is strongly suppressed, but they can form in early
Universe and strongly influence cosmological evolution of 4th
generation \index{4th
generation} hadrons. As shown in \cite{I}, \underline{an}ti-
\underline{U}-\underline{t}riple state called \underline{anut}ium
or $\Delta^{--}_{3 \bar U}$ is of special interest. This stable
anti-$\Delta$-isobar, composed of $\bar U$ antiquarks is bound with $^4He$ in atom-like state
of O-helium \index{O-helium}\cite{I} or ANO-helium \index{ANO-helium}\cite{lom,KPS06,Khlopov:2006dk},
proposed as dominant forms of dark matter \index{dark matter}.

\subsection{Stable Techniparticles}

The minimal walking technicolor \index{technicolor} model
\cite{Sannino:2004qp,Hong:2004td,Dietrich:2005jn,Dietrich:2005wk,Gudnason:2006ug,Gudnason:2006yj}
has two techniquarks, i.e. up $U$ and down $D$, that transform
under the adjoint representation of an $SU(2)$ technicolor gauge
group. The global symmetry of the model is an $SU(4)$ that breaks
spontaneously to an $SO(4)$. The chiral condensate of the
techniquarks breaks the electroweak symmetry. There are nine
Goldstone bosons emerging from the symmetry breaking. Three of
them are eaten by the $W$ and the $Z$ bosons. The remaining six
Goldstone bosons are $UU$, $UD$, $DD$ and their corresponding
antiparticles. These Goldstone bosons carry technibaryon number since they are made of
two techniquarks or two anti-techniquarks. It means that if no
processes violate the technibaryon \index{technibaryon} number, the lightest
technibaryon will be stable. The electric charges of $UU$, $UD$,
and $DD$ are given in general by $y+1$, $y$, and $y-1$
respectively, where $y$ is an arbitrary real number. For any real
value of $y$, gauge anomalies are
cancelled~\cite{Gudnason:2006yj}. The model requires in addition
the existence of a fourth family of leptons, i.e. a ``new
neutrino'' $\nu'$ and a ``new electron'' $\zeta$ in order to
cancel the Witten global anomaly. Their electric charges are in
terms of $y$ respectively $(1-3y)/2$ and $(-1-3y)/2$. The
effective theory of this minimal walking technicolor model has
been presented in~\cite{Gudnason:2006ug,Foadi:2007ue}.

Scenario of composite dark matter \index{dark matter} \cite{KK} corresponds to the choice $y=1$, when Goldstone bosons $UU$,
$UD$, and $DD$ have electric charges 2, 1, and 0 respectively. In
addition for $y=1$, the electric charges of $\nu'$ and $\zeta$ are
respectively $-1$ and $-2$. There are three possibilities for a scenario where
stable particles with $-2$ electric charge have substantial relic
densities and can capture $^4He^{++}$ nuclei to form a neutral
atom. The first
one is to have an excess of $\bar{U}\bar{U}$, which has $-2$
charge. For this to be true one should assume that $UU$ is lighter
than $UD$ and $DD$ and no processes (apart from electroweak
sphalerons) violate the technibaryon \index{technibaryon} number $TB$. The second one is to
have excess of $\zeta$ that has $-2$ charge, being lighter than $\nu'$, and
there should be no mixing between the fourth family of leptons and
the other three of the Standard Model. The technilepton \index{technilepton} number $L'$ is violated
only by sphalerons and therefore after the
sphalerons freeze out, $L'$ is conserved, which means that the
lightest particle, that is $\zeta$ in this case, is absolutely
stable. Finally there is a
possibility to have both $L'$ and $TB$ conserved after sphalerons
have frozen out. In this case, both $\bar{U}\bar{U}$ and $\zeta$ take part 
in formation of composite dark matter.

\section{Composite dark matter}
\subsection{Problems of composite dark matter}
Glashow's sinister Universe was first inspiring example of
composite dark matter scenario. The problem of such scenario is
inevitable presence of "products of incomplete combustion" and the
necessity to decrease their abundance. Indeed the
tera-lepton and tera-hadron relics from intermediate stages of a
multi-step process towards a final $(UUUEE)$ formation must
survive with high abundance of {\it visible} relics in the present
Universe. 

The model \cite{Glashow} didn't offer any physical mechanism for
generation of cosmological tera-particle asymmetry and such
asymmetry was postulated to saturate the observed dark matter \index{dark matter}
density. This assumption was taken in \cite{Fargion:2005xz} and it
was revealed that while the assumed tera-baryon asymmetry for $U$
washes out by annihilation primordial $\bar U$, the tera-lepton
asymmetry of $E^-$ can not effectively suppress the abundance of
tera-positrons $E^+$ in the earliest as well as in the late
Universe stages. 
Moreover ordinary $^4$He formed in Standard Big Bang
Nucleosynthesis \index{Standard Big Bang
Nucleosynthesis} binds at $T \sim 15 keV$ virtually all the free
$E^-$ into positively charged $(^4HeE^-)^+$ "ion", which puts
Coulomb barrier for any successive $E^-E^+$ annihilation or any
effective $EU$ binding.

The remaining abundance of $(eE^+)$ and $(^4HeE^-e)$ exceeds by
{\it 27 orders} of magnitude the terrestrial upper limit for
anomalous hydrogen and it can not be removed by any mechanism \cite{Fargion:2005xz}. Therefore, while tera helium $(UUUEE)$
would co-exist with observational data, being an interesting
candidate for dark matter \index{dark matter}, its tera-lepton partners poison and
forbid this opportunity.

The approaches (b) and (c) try to escape the problems of free
charged dark matter particles \cite{Dimopoulos:1989hk} by hiding
opposite-charged particles in atom-like bound systems, which
interact weakly with baryonic matter. However, in the case of
charge symmetry, when primordial abundances of particles and
antiparticles are equal, annihilation in the early Universe
suppresses their concentration. If this primordial abundance still
permits these particles and antiparticles to be the dominant dark
matter, the explosive nature of such dark matter is ruled out by
constraints on the products of annihilation in the modern Universe
\cite{Q,FKS}. Even in the case of charge asymmetry with primordial
particle excess, when there is no annihilation in the modern
Universe, binding of positive and negative charge particles is
never complete and positively charged heavy species should retain.
Recombining with ordinary electrons, these heavy positive species
give rise to cosmological abundance of anomalous isotopes,
exceeding experimental upper limits. To satisfy these upper
limits, the anomalous isotope \index{anomalous isotope} abundance on Earth should be
reduced, and the mechanisms for such a reduction are accompanied
by effects of energy release which are strongly constrained, in
particular, by the data from large volume detectors.

These problems of composite dark matter \index{dark matter} models \cite{Glashow,5}
revealed in \cite{Q,Fargion:2005xz,FKS,I}, can be avoided, if the
excess of only $-2$ charge $A^{--}$
particles is generated in the early Universe. In
walking technicolor \index{technicolor} models, technilepton \index{technilepton} and technibaryon \index{technibaryon} excess
is related to cosmological baryon asymmetry and the excess of $-2$ charged
particles can appear naturally for a reasonable choice of model
parameters \cite{KK}. It distinguishes this case from other composite dark
matter models (see \cite{Khlopov:2006dk} for their review), since in all the previous realizations, starting
from \cite{Glashow}, such an excess was put by hand to saturate
the observed dark matter density by composite dark
matter. On that reason we'll further follow scenario \cite{KK}, 
based on walking technicolor model.

 \subsection{\label{Formation} Formation of O-helium}
In the Big Bang nucleosynthesis, $^4He$ is formed with an
abundance $r_{He} = 0.1 r_B = 8 \cdot 10^{-12}$ and, being in
excess, binds all the negatively charged techni-species into
atom-like systems.

At a temperature $T<I_{o} = Z_{TC}^2 Z_{He}^2 \alpha^2 m_{He}/2
\approx 1.6 \MeV,$ where $\alpha$ is the fine structure constant,
and $Z_{TC}=-2$ stands for the electric charge of $\bar{U}\bar{U}$
and/or of $\zeta$, the reaction \beq
\zeta^{--}+^4He^{++}\rightarrow \gamma +(^4He\zeta) \label{EHeIg}
\eeq and/or \beq (\bar{U}\bar{U})^{--}+^4He^{++}\rightarrow \gamma
+(^4He(\bar{U}\bar{U})) \label{EHeIgU} \eeq can take place. In
these reactions neutral techni-O-helium \index{techni-O-helium} ``atoms" ($tOHe$) are produced. The
size of these ``atoms" is \cite{I,FKS} \beq R_{o} \sim 1/(Z_{TC}
Z_{He}\alpha m_{He}) \approx 2 \cdot 10^{-13} \cm. \label{REHe}
\eeq
Virtually all the free $(\bar{U}\bar{U})$ and/or $\zeta$ (which will be further denoted by
$A^{--}$) are trapped by helium and their remaining abundance
becomes exponentially small.

For particles $Q^-$ with charge $-1$, as for tera-electrons in the
sinister model \cite{Glashow}, $^4He$ trapping results in the
formation of a positively charged ion $(^4He^{++} Q^-)^+$, result
in dramatic over-production of anomalous hydrogen
\cite{Fargion:2005xz}. Therefore, only the choice of $- 2$
electric charge for stable techniparticles makes it possible to
avoid this problem. In this case, $^4He$ trapping leads to the
formation of neutral techni-O-helium {\it techni-O-helium} ``atoms"
$(^4He^{++}A^{--})$.

This is the case for all the realistic models of composite dark matter, involving -2 charged particles.
Binding of various types of such particles with $^4He$ results in different forms of O-helium atoms: O-helium \index{O-helium} \cite{I}, OLe-helium \index{OLe-helium} \cite{FKS}, ANO-helium \index{ANO-helium} \cite{KPS06,Khlopov:2006dk} or techni-O-helium \index{techni-O-helium} 
\cite{KK}. However, all these different forms of O-helium have the same size (\ref{REHe}), the same cross section for interaction with baryonic matter and play the same
role in nuclear transformations. Therefore in the successive discussion we'll refer to general properties of O-helium
and specify the particular case of techni-O-helium, when specific features of walking technicolor scenario are considered.
 
The formation of techni-O-helium \index{techni-O-helium} reserves a fraction of $^4He$ and
thus it changes the primordial abundance of $^4He$. For the
lightest possible masses of the techniparticles $m_{\zeta} \sim
m_{TB} \sim 100 \GeV$, this effect can reach 50\% of the $^4He$
abundance formed in SBBN. Even if the mass of the techniparticles
is of the order of TeV, $5\%$ of the $^4He$ abundance is hidden in
the techni-O-helium atoms. This can lead to important consequences
once we compare the SBBN theoretical predictions to observations.
\subsection{Primordial heavy elements from O-helium catalysis}
The question of the participation of O-helium \index{O-helium} in nuclear
transformations and its direct influence on the chemical element
production is less evident. Indeed, O-helium looks like an
$\alpha$ particle with a shielded electric charge. It can closely
approach nuclei due to the absence of a Coulomb barrier. Because
of this in the presence of O-helium the
character of SBBN processes should change drastically. However, it
might not lead to immediate contradictions with the observational data.

The following simple argument \cite{FKS,KK} can be used for suppression of  
binding of $A^{--}$ with nuclei heavier than $^4He$
in nuclear transformations \index{nuclear transformations}. In fact,
the size of O-helium is of the order of the size of $^4He$
and for a nucleus $^A_ZQ$ with electric charge $Z>2$, the size of
the Bohr orbit for an $Q A^{--}$ ion is less than the size of the
nucleus $^A_ZQ$ (see \cite{Cahn}). This means that while binding with a heavy
nucleus, $A^{--}$ penetrates it and interacts effectively with a
part of the nucleus of a size less than the corresponding Bohr
orbit. This size corresponds to the size of $^4He$, making
O-helium the most bound $Q A^{--}$ atomic state. It favors
a picture, according to which a O-helium collision with a
nucleus, results in the formation of O-helium \index{O-helium} and the whole
process looks like an elastic collision.

The interaction of the $^4He$ component of $(He^{++}A^{--})$ with
a $^A_ZQ$ nucleus can lead to a nuclear transformation due to the
reaction \beq ^A_ZQ+(HeA) \rightarrow ^{A+4}_{Z+2}Q +A^{--},
\label{EHeAZ} \eeq provided that the masses of the initial and
final nuclei satisfy the energy condition \beq M(A,Z) + M(4,2) -
I_{o}> M(A+4,Z+2), \label{MEHeAZ} \eeq where $I_{o} = 1.6 \MeV$ is
the binding energy \index{binding energy} of O-helium and $M(4,2)$ is the mass of
the $^4He$ nucleus.

This condition is not valid for stable nuclei participating in
reactions of the SBBN. However, tritium $^3H$, which is also
formed in SBBN with abundance $^3H/H \sim 10^{-7}$ satisfies this
condition and can react with O-helium, forming $^7Li$ and
opening the path of successive O-helium catalyzed
transformations to heavy nuclei. This effect might strongly
influence the chemical evolution of matter on the pre-galactic
stage and needs a self-consistent consideration within the Big
Bang nucleosynthesis network. However, the following arguments \cite{FKS,KK}
show that this effect may not lead to immediate contradiction with
observations as it might be expected.
\begin{itemize}
\item[$\bullet$] On the path of reactions (\ref{EHeAZ}), the final
nucleus can be formed in the excited $(\alpha, M(A,Z))$ state,
which can rapidly experience an $\alpha$- decay, giving rise to
O-helium regeneration and to an effective quasi-elastic
process of $(^4He^{++}A^{--})$-nucleus scattering. It leads to a
possible suppression of the O-helium catalysis of nuclear
transformations \index{nuclear
transformations}. \item[$\bullet$] The path of reactions
(\ref{EHeAZ}) does not stop on $^7Li$ but goes further through
$^{11}B$, $^{15}N$, $^{19}F$, ... along the table of the chemical
elements. \item[$\bullet$] The cross section of reactions
(\ref{EHeAZ}) grows with the mass of the nucleus, making the
formation of the heavier elements more probable and moving the
main output away from a potentially dangerous Li and B
overproduction.
\end{itemize}
Such a qualitative change of the
physical picture appeals to necessity in a detailed nuclear
physics treatment of the ($A^{--}$+ nucleus) systems and of the
whole set of transformations induced by O-helium. Though the above arguments
do not seem to make these dangers immediate and obvious, a
detailed study of this complicated problem is needed.

The first publications on possible realistic composite dark matter \index{dark matter}
scenarios \cite{I,FKS} gave rise to the development of another
aspect of the problem, the Charged massive particles BBN (CBBN),
studying the influence of {\it unstable} negatively charged massive
particles on BBN
\cite{Pospelov:2006sc,Kaplinghat,Kohri:2006cn,Kino,Kawasaki:2007xb,Jedamzik:2007cp,Pospelov:2007js}.
Bound states of metastable singly charged particle $X^-$ with nuclei can catalyze formation of lithium and even
elements with $A>8$ (e.g. $^9Be$ \cite{Pospelov:2007js})
The important difference of CBBN \index{CBBN} considered in these papers, from
our approach, is that singly charged particles $X^-$ with charge
$-1$ do not screen the $+2$ charge of $He$ in a $(HeX)^+$ ion-like
bound system, and the Coulomb barrier of the $(HeX)^+$ ion can
strongly hamper the path for the creation of heavy isotopes.

It should be also noted that the self-consistent picture of CBBN inevitably involves
analysis of non-equilibrium nucleosynthesis reactions, induced by
hadronic and electromagnetic cascades from products of $X$ decays.
The latter effect, which was discussed in \cite{Kohri:2006cn},
implies the theory of
non-equilibrium cosmological nucleosynthesis
\cite{Linde3,Linde4,book} (see also
\cite{Dimopoulos:1987fz,Kawasaki:2004yh,Kawasaki:2004qu,Kohri:2005wn,Jedamzik:2006xz}).

\section{\label{Interactions} Tests for O-helium and its -2 charge constituents}

\subsection{Techniparticle component of cosmic rays}
 The nuclear interaction of techni-O-helium with cosmic rays \index{cosmic rays} gives rise to
ionization of this bound state in the interstellar gas and to
acceleration of free techniparticles \index{techniparticles} in the Galaxy. During the
lifetime of the Galaxy $t_G \approx 3 \cdot 10^{17} \s$, the
integral flux of cosmic rays $$F(E>E_0)\approx 1 \cdot
\left(\frac{E_0}{1 \GeV} \right)^{-1.7} \cm^{-2} \s^{-1}$$ can
disrupt the fraction of galactic techni-O-helium \index{techni-O-helium} $ \sim
F(E>E_{min}) \sigma_o t_G \le 10^{-3},$ where we took $E_{min}
\sim I_o.$ Assuming a universal mechanism of cosmic ray
acceleration \index{cosmic ray
acceleration}, a universal form of their spectrum, taking into
account that the $^4He$ component corresponds to $\sim 5$\% of the
proton spectrum, and that the spectrum is usually reduced to the
energy per nucleon, the $-2$ charged
techniparticle component with anomalous low $Z/A$ can be present in cosmic rays \index{cosmic rays} at a level
of \cite{KK}\beq
 \frac{A^{--}}{He} \ge 3 \cdot 10^{-7} \cdot S_2^{-3.7}, \eeq
 where $S_2=(m_o/100\GeV)$ and $m_o$ is the mass of O-helium \index{O-helium}.
 This flux may be within the reach for PAMELA and AMS02 cosmic ray
experiments.

Recombination of free techniparticles with protons and nuclei in
the interstellar space can give rise to radiation in the range
from few tens of keV - 1 MeV. However such a radiation is below
the cosmic nonthermal electromagnetic background radiation
observed in this range  \cite{FKS,KK}.

\subsection{\label{EpMeffects} O-helium catalyzed processes in the Earth}
The first evident consequence of the proposed model is an
inevitable presence of $tOHe$ in terrestrial matter. This is
because terrestrial matter appears opaque to O-helium \index{O-helium} and stores all
its in-falling flux.

If the $tOHe$ capture by nuclei is not effective, its diffusion in
matter is determined by elastic collisions, which have a transport
cross section per nucleon \beq \sigma_{tr} = \pi R_{o}^2
\frac{m_{p}}{m_o} \approx 10^{-27}/S_2 \cm^{2}. \label{sigpEpcap}
\eeq In atmosphere, with effective height $L_{atm} =10^6 \cm$ and
baryon number density $n_B= 6 \cdot 10^{20} \cm^{-3}$, the opacity
condition $n_B \sigma_{tr} L_{atm} = 6 \cdot 10^{-1}/S_2$ is not
strong enough. Therefore, the in-falling $tOHe$ particles are
effectively slowed down only after they fall down terrestrial
surface in $16 S_2$ meters of water (or $4 S_2$ meters of rock).
Then they drift with velocity $V = \frac{g}{n \sigma v} \approx 8
S_2 A^{1/2} \cm/\s$ (where $A \sim 30$ is the average atomic
weight in terrestrial surface matter, and $g=980~
\cm/\s^2$), sinking down the center of the Earth on a
timescale $t = R_E/V \approx 1.5 \cdot 10^7 S_2^{-1}\s$, where
$R_E$ is the radius of the Earth.

The in-falling techni-O-helium \index{techni-O-helium} flux from dark matter halo is
$\mathcal{F}=n_o v_h/8 \pi$, where the number density of $tOHe$ in
the vicinity of the Solar System is $n_o=3 \cdot 10^{-3}S_2^{-1}
\cm^{-3}$ and the averaged velocity $v_h \approx 3 \cdot
10^{7}\cm/\s$. During the lifetime of the Earth ($t_E \approx
10^{17} \s$), about $2 \cdot 10^{38}S_2^{-1}$ O-helium
atoms were captured. If $tOHe$ dominantly sinks down the Earth, it
should be concentrated near the Earth's center within a radius
$R_{oc} \sim \sqrt{3T_c/(m_o 4 \pi G \rho_c)}$, which is $\le 10^8
S_2^{-1/2}\cm$, for the Earth's central temperature $T_c \sim 10^4
\K$ and density $\rho_c \sim 4 \g/\cm^{3}$.

Near the Earth's surface, the O-helium \index{O-helium} abundance is
determined by the equilibrium between the in-falling and
down-drifting fluxes. It gives $$n_{oE}=2 \pi \mathcal{F}/V = 3
\cdot 10^3 \cdot S_2^{-2}\cdot A^{-1/2}\cm^{-3},$$ or for $A \sim
30$ about $5 \cdot 10^2 \cdot S_2^{-2} \cm^{-3}$. This number
density corresponds to the fraction $$f_{oE} \sim 5 \cdot 10^{-21}
\cdot S_2^{-2}$$ relative to the number density of the terrestrial
atoms $n_A \approx 10^{23} \cm^{-3}$.

These neutral $(^4He^{++}A^{--})$ ``atoms" may provide a catalysis
of cold nuclear reactions in ordinary matter (much more
effectively than muon catalysis). This effect needs a special and
thorough investigation. On the other hand, if $A^{--}$ capture by
nuclei, heavier than helium, is not effective and does not lead to
a copious production of anomalous isotopes \index{anomalous isotopes}, the
$(^4He^{++}A^{--})$ diffusion in matter is determined by the
elastic collision cross section (\ref{sigpEpcap}) and may
effectively hide O-helium from observations.

One can give the following argument for an effective regeneration
and quasi-elastic collisions of O-helium \index{O-helium} in terrestrial
matter. The O-helium can be destroyed in the reactions
(\ref{EHeAZ}). Then, free $A^{--}$ are released and due to a
hybrid Auger effect (capture of $A^{--}$, ejection of ordinary $e$
from the atom with atomic number $A$, and charge of the nucleus
$Z$), $A^{--}$-atoms are formed, in which $A^{--}$ occupies highly
an excited level of the $(_Z^AQA)$ system, which is still much
deeper than the lowest electronic shell of the considered atom.
The $(_Z^AQA)$ atomic transitions to lower-lying states cause
radiation in the intermediate range between atomic and nuclear
transitions. In course of this falling down to the center of the
$(Z-A^{--})$ system, the nucleus approaches $A^{--}$. For $A>3$
the energy of the lowest state $n$ (given by $E_n=\frac{M \bar
\alpha^2}{2 n^2} = \frac{2 A m_p Z^2 \alpha^2}{n^2}$)  of the
$(Z-A^{--})$ system (having reduced mass $M \approx A m_p$) with a
Bohr orbit $r_n=\frac{n}{M \bar \alpha}= \frac{n}{2 A Z m_p
\alpha}$, exceeding the size of the nucleus $r_A \sim
A^{1/3}m^{-1}_{\pi}$ ($m_{\pi}$ being the mass of the pion), is
less than the binding energy of $tOHe$. Therefore the regeneration
of O-helium in a reaction, inverse to (\ref{EHeAZ}), takes
place. An additional reason for the domination of the elastic
channel of the reactions (\ref{EHeAZ}) is that the final state
nucleus is created in the excited state and its de-excitation via
$\alpha$-decay can also result in O-helium \index{O-helium} regeneration. If
regeneration is not effective and $A^{--}$ remains bound to the
heavy nucleus, anomalous isotope of $Z-2$ element should appear.
This is a serious problem for the considered model.

However, if the general picture of sinking down is valid, it might
give no more than the ratio $f_{oE} \sim 5 \cdot 10^{-21} \cdot S_2^{-2}$
of number density of anomalous isotopes to the number density of atoms of
terrestrial matter around us, which is below the experimental
upper limits for elements with $Z \ge 2$. For comparison, the best
upper limits on the anomalous helium were obtained in \cite{exp3}.
It was found, by searching with the use of laser spectroscopy for
a heavy helium isotope in the Earth's atmosphere, that in the mass
range 5 GeV - 10000 GeV, the terrestrial abundance (the ratio of
anomalous helium number to the total number of atoms in the Earth)
of anomalous helium is less than $2 \cdot 10^{-19}$ - $3 \cdot
10^{-19}$.

\subsection{\label{DMdirect} Direct search for O-helium}

It should be noted that the nuclear cross section of the O-helium
\index{O-helium} interaction with matter escapes the severe
constraints \cite{McGuire:2001qj} on strongly interacting dark
matter \index{dark
matter} particles (SIMPs) \cite{Starkman,McGuire:2001qj} imposed by
the XQC experiment \cite{XQC}.

In underground detectors, $tOHe$ ``atoms'' are slowed down to
thermal energies and give rise to energy transfer $\sim 2.5 \cdot
10^{-3} \eV A/S_2$, far below the threshold for direct dark matter
detection. It makes this form of dark matter \index{dark matter} insensitive to the
CDMS constraints. However, $tOHe$ induced nuclear transformation \index{nuclear transformation}
can result in observable effects.

 Therefore, a special strategy of such a search is needed, that
can exploit sensitive dark matter \index{dark matter} detectors
 on the ground or in space. In particular, as it was revealed in
\cite{Belotsky:2006fa}, a few $\g$ of superfluid $^3He$ detector
\cite{Winkelmann:2005un}, situated in ground-based laboratory can
be used to put constraints on the in-falling O-helium flux
from the galactic halo.
\section{\label{Discussion} Discussion}
To conclude, the existence of heavy stable particles can
offer new solutions for dark matter \index{dark matter} problem. 
If stable particles have electric charge, dark matter candidates
can be atom-like states, in which negatively and positively 
charged particles are bound by Coulomb attraction. In this case
there is a serious problem to prevent overproduction of accompanying anomalous forms of
atomic matter.

Indeed, recombination of charged species is never complete in the
expanding Universe, and significant fraction of free charged
particles should remain unbound. Free positively charged species
behave as nuclei of anomalous isotopes \index{anomalous isotopes}, giving rise to a danger of
their over-production. Moreover, as soon as $^4He$ is formed in
Big Bang nucleosynthesis it captures all the free negatively
charged heavy particles. If the charge of such particles is -1 (as
it is the case for tera-electron in \cite{Glashow}) positively
charged ion $(^4He^{++}E^{-})^+$ puts Coulomb barrier for any
successive decrease of abundance of species, over-polluting modern
Universe by anomalous isotopes. It excludes the possibility of
composite dark matter with $-1$ charged constituents and only $-2$
charged constituents avoid these troubles, being trapped by helium
in neutral OLe-helium \index{OLe-helium}, O-helium \index{O-helium} (ANO-helium\index{ANO-helium}) or  techni-O-helium \index{techni-O-helium} states.

The existence of $-2$ charged states and the absence of stable
$-1$ charged constituents can take place in AC-model \index{AC-model}, in charge
asymmetric model of 4th generation \index{4th generation} and in
 walking technicolor \index{technicolor} model with stable doubly charged technibaryons \index{technibaryons} and/or technileptons
\index{technileptons}. 

To avoid overproduction of anomalous isotopes\index{anomalous isotopes}, an excess of $-2$
charged particles over their antiparticles in the early Universe is sufficient. In the earlier realizations of
composite dark matter scenario, this excess was put by hand to
saturate the observed dark matter\index{dark matter} density. In walking technicolor\index{technicolor} model this
abundance of -2 charged techibaryons\index{techibaryons} and/or technileptons\index{technileptons} is connected
naturally to the baryon relic density. These doubly charged $A^{--}$
techniparticles bind with $^4He$ in the techni-O-helium
 \index{techni-O-helium} neutral states. For reasonable
values of the techniparticle mass, the amount of primordial $^4He$,
bound in this atom like state is significant and should be taken
into account in comparison to observations.

A challenging problem is the nuclear
transformations \index{nuclear
transformations}, catalyzed by O-helium\index{O-helium}. The question about
their consistency with observations remains open, since special
nuclear physics analysis is needed to reveal what are the actual
O-helium effects in SBBN and in terrestrial matter. However, qualitatively
one can expect much easier path for O-helium catalysis of primordial heavy elements \index{primordial heavy elements},
than in CBBN \index{CBBN} \cite{Pospelov:2006sc,Kaplinghat,Kohri:2006cn,Kino,Kawasaki:2007xb,Jedamzik:2007cp,Pospelov:2007js}
or in nonthermal nucleosynthesis \index{nonthermal nucleosynthesis} \cite{Linde3,Linde4,book,Dimopoulos:1987fz,Kawasaki:2004yh,Kawasaki:2004qu,Kohri:2005wn,Jedamzik:2006xz}).

The destruction of O-helium\index{O-helium} by cosmic rays in the Galaxy
releases free charged heavy stable particles particles, which can be accelerated
and contribute to the flux of cosmic rays. In this context, the
search for stable charged particles in cosmic rays\index{cosmic rays} and at accelerators
acquires the meaning of a crucial test for existence of 
basic constituents of composite dark matter\index{dark matter}. 

Models of composite dark matter enrich the class of possible stable particles, which can follow from
extensions of the Standard Model and be considered as dark matter candidates.
One can extend the generally accepted viewpoint that new stable particles should be neutral and weakly interacting
as follows: they can also be charged and play the role of DARK matter because they are hidden in 
atom-like states, which are not the source of visible light. Formation of O-helium\index{O-helium} and nuclear transformations\index{nuclear transformations}, catalyzed by it, are inevitable consequences of this extension.  
It makes the existence of pregalactic heavy elements (like carbon, nitrogen, oxygen, neon etc)
a signature for composite dark matter\index{dark matter}. Astronomical observations might favour this prediction.
Still no astronomical objects are observed without heavy elements.

\section*{Acknowledgements}
 I am grateful to Organizers of Blois2007 Conference for creative atmosphere of fruitful discussions.


\end{document}